\documentclass[psamsfonts]{amsart}
\usepackage{hyperref}
\usepackage{amsthm,amssymb}
\usepackage{enumerate}
\usepackage[english]{babel}
\newcommand{\RR}{\mathbb{R}}
\newcommand{\CC}{\mathbb{C}}
\newcommand{\NN}{\mathbb{N}}
\newcommand{\ZZ}{\mathbb{Z}}

\newcommand{\PP}{\mathbb{P}}

\newcommand{\colouring}{\varLambda}

\newcommand{\cB}{\mathcal{B}}
\newcommand{\cC}{\mathcal{C}}

\newcommand{\cG}{\mathcal{G}}
\newcommand{\cH}{\mathcal{H}}

\newcommand{\cL}{\mathcal{L}}

\newcommand{\lr}{\mathcal{R}} 
\newcommand{\calh}{\mathcal{H}}

\newcommand{\dd}{\mathrm{d}}
\newcommand{\Tr}{{\mathop{\mathrm{Tr}}}}

\DeclareMathOperator{\supp}{\mathrm{supp}}

\newcommand{\Gammaf}{\Gamma_{\mbox{full}}}

\newcommand{\init}{\iota} 
\newcommand{\term}{\tau} 
\newcommand{\cals}{\mathcal{F}} 
\newcommand{\cala}{\mathcal{A}} 
\newcommand{\calpM}{\mathcal{P}_0^B (M)} 
\newcommand{\domain}{\mathcal{D}}
\newcommand{\fundom}{\mathfrak{D}}
\newcommand{\rightcont}{\mathcal{R}}
\newcommand{\Gammat}{\Gamma_{\mbox{trans}}}
\newcommand{\vispt}{\mathcal{VP}}
\newcommand{\elzh}{\ell^2 (\ZZ^d,\cH)}

\newcommand{\be}{\begin{equation}}
\newcommand{\ee}{\end{equation}}
\newcommand{\bea}{\begin{eqnarray*}}
\newcommand{\eea}{\end{eqnarray*}}

\newtheorem{thm}{Theorem}
\newtheorem{lem}[thm]{Lemma}
\newtheorem{prp}[thm]{Proposition}
\newtheorem{cor}[thm]{Corollary}
\theoremstyle{definition}
\newtheorem{dfn}[thm]{Definition}
\newtheorem{ass}{Assumption}
\theoremstyle{remark}
\newtheorem{rem}[thm]{Remark}

\newcommand{\Hm}[1]{\leavevmode{\marginpar{\tiny%
$\hbox to 0mm{\hspace*{-0.5mm}$\leftarrow$\hss}%
\vcenter{\vrule depth 0.1mm height 0.1mm width \the\marginparwidth}%
\hbox to
0mm{\hss$\rightarrow$\hspace*{-0.5mm}}$\\\relax\raggedright #1}}}

\usepackage[usenames]{color}

\begin{document}
\title[Uniform existence of the IDS]{Uniform existence of the integrated density of states for combinatorial and metric graphs over $\ZZ^d$}

\author[M.~J.~Gruber]{Michael J.\ Gruber}
\address[M.G.]{TU Clausthal\\
Institut f\"ur Mathematik\\
38678 Clausthal-Zellerfeld\\
Germany}
\urladdr{\url{http://www.math.tu-clausthal.de/~mjg/}}

\author[D.~H.~Lenz]{Daniel H.\ Lenz}
\address[D.L.]{Fakult\"at f\"ur Mathematik, TU Chemnitz, 09107 Chemnitz, Germany}
\urladdr{\url{http://www.tu-chemnitz.de/mathematik/analysis/dlenz}}
\curraddr[D.L.]{Department of Mathematics \\ Rice University\\ Houston TX 77005-1892\\ USA}

\author[I.~Veseli\'c]{Ivan Veseli\'c}
\address[I.V.]{Emmy-Noether-Programm der Deutschen Forschungsgemeinschaft \&
Fakult\"at f\"ur Mathematik, TU Chemnitz, 09107 Chemnitz, Germany}
\urladdr{\url{http://www.tu-chemnitz.de/mathematik/enp}}
\curraddr[I.V.]{Institut f\"ur Angewandte Mathematik\\ 53115 Universit\"at Bonn\\ Germany}

\keywords{random Schr\"odinger operator, combinatorial graph, metric graph, quantum graph, integrated density of states} 
\subjclass[2000]{Primary 47E05; Secondary 34L40, 47B80, 47N50, 60H25, 81Q10}
\copyrightinfo{2007}{by the authors. Faithful reproduction of this article
in its entirety is permitted for non-commercial purposes.}

\begin{abstract}
We give an overview and extension of recent results on
ergodic random Schr\"odinger operators for models on $\ZZ^d$.
The operators we consider are defined on combinatorial or metric graphs,
 with random potentials, random boundary conditions and random metrics
taking values in a finite set.  
We show that normalized finite volume eigenvalue counting functions converge to a limit uniformly in the
energy variable, at least locally.
This limit, the integrated density of states (IDS), can be expressed by a closed Shubin-Pastur type trace formula.  
The set of points of increase of the IDS supports the spectrum and its points of discontinuity are characterized by
existence of compactly supported eigenfunctions. This applies to several examples, including
various periodic operators and percolation models.
\end{abstract}

\maketitle
\let\languagename\relax

\bigskip

\section{Introduction}
This paper deals with  spectral analysis of certain random type  operators on  graphs with a $\ZZ^d$-structure. We consider both combinatorial graphs and quantum graphs.  Randomness enters not only  via potentials  but, more importantly, via geometry. More precisely, we will consider certain  random ``perturbations'' of our graph  in the combinatorial setting and random boundary conditions and random lengths in the quantum graph setting.   Our results are concerned with existence of the integrated density of states for such models.  

The   \textit{integrated density of states} (IDS) or \textit{spectral distribution function}  is a fundamental tool in the study of  such random operators. It measures  the number of states (up to a given energy) per unit volume of the underlying system. Accordingly, it can be obtained as a limit of normalized eigenvalue counting functions.   Existence of this  limit in the sense of pointwise convergence or rather vague convergence of measures is well established for ergodic Schr\"odinger operators in the continuum (i.e.~on $L^2(\RR^d)$)
and on the lattice (i.e.~on $\ell^2(\ZZ^d)$), 
see for instance the early papers \cite{Pastur-80,Shubin-82,KirschM-82c}
and the recent surveys \cite{KirschM-07,Veselic-07b}. Some of these approaches 
can be modified to give an analogous result for random Schr\"odinger on metric graphs \cite[\S~6]{HelmV}.

It turns out that existence of the limit can be shown in a much stronger sense than in the pointwise one, 
viz in the sense of convergence in the supremum norm. Thus, the limit exists uniformly in the energy. This is particularly remarkable  as the limiting object, the IDS, can have many points of discontinuity
in the setting of geometric randomness described above.   As such  discontinuities are not possible for random Schr\"odinger operators on $\ZZ^d$ itself, they are an exclusive consequence of geometric ingredients.  In fact, it turns out that they are intimately related to  local features of the geometry viz to existence of compactly supported eigenfunctions.  
This phenomenon has attracted attention before.  For  periodic, abelian graphs it was observed by Kuchment in \cite{Kuchment-91,Kuchment-05} (see \cite{DodziukLMSY-03} for related material as well). 
For discrete models it was  then  studied systematically  
in the somewhat different context of aperiodic order in \cite{KlassertLS-03}, as well as for 
periodic and percolation models on quasi-transitive graphs in \cite{Veselic-05b}. These two types 
of results were unified in the recent \cite{LenzV}.
Random operators on graphs based on $\ZZ^d$ allow for a particularly nice 
treatment of the question of uniform convergence, see \cite{LenzMV,GruberLV-07}.

This paper deals with the circle of topics just discussed for models 
with the abovementioned $\ZZ^d$-structure. 
In particular, it  surveys and extends the results of \cite{LenzMV, GruberLV-07}.  Examples  include several variants of periodic operators and percolation models. One essential tool in these works is a general ergodic type result from \cite{LenzMV}.  This result is not about operators, but about Banach space valued functions which are compatible with a so-called colouring. In order to apply it, we need a certain finiteness assumption to hold for the number of local geometric situations in our model.  Limitations and extensions of this type of approach are discussed in the final section. 

More generally, the paper is organized as follows.
In Section \ref{s-ET} we define colourings of $\ZZ^d$ and present the mentioned general ergodic theorem from \cite{LenzMV}, as well as a random version.
In Sections \ref{s-comb-graphs}, \ref{s-metric-graphs} and \ref{s-rndlen} 
we apply these results to operators on combinatorial graphs, metric graphs and metric graphs
with random lengths, respectively. This corresponds to three different types of behaviour
for the underlying (counting) functions and
allows us to cover examples ranging from periodic operator to random order,
including aperiodic order and percolation.  
Finally, Section~\ref{s-jumps} gives an outlook on the interplay between uniform convergence
and discontinuities of the integrated density of states.

The results in Section \ref{s-comb-graphs} are taken from \cite{LenzMV}. 
The results in Section \ref{s-metric-graphs} are taken from \cite{GruberLV-07}. The results in Section \ref{s-rndlen} are new.

\section{Colourings and Ergodic Theorems}\label{s-ET}
In this section, we introduce some basic set-up and  discuss the abstract ergodic theorem of \cite{LenzMV} (see \cite{Lenz-02, LenzS-06} for earlier  results of the same type).  The theorem is phrased in terms of   Banach-space valued functions on patterns. It will turn out that eigenvalue counting functions of suitable (restrictions of) operators provide exactly such Banach-space valued functions. 

\medskip

We are concerned with the graph $\ZZ^d$. Thus, the vertex set is given by $\ZZ^d$ and  vertices of Euclidean distance $1$ are adjacent.
There are two groups acting naturally on  this graph. One is the group $\Gammat=\ZZ^d$ acting by translations. The other group is the group $\Gammaf$ of all graph automorphisms.  Note that $\Gammaf$ is generated by translations by
vectors in $\ZZ^d$ and a finite set of rotations.  For our results it does not matter which of the two groups  we consider.  Thus, we now choose  $\Gamma$
to be $\Gammaf$ or $\Gammat$.
  The action will be written multiplicatively. 

Let $\cala$ be a finite set. The set of all
finite subsets of $\ZZ^d$ is denoted by $\cals$.   A map $\colouring \colon
\ZZ^d \longrightarrow \cala$ is called an \emph{$\cala$-colouring} of
$\ZZ^d$. A map $ P \colon Q(P) \longrightarrow \cala$ with $Q(P)\in
\cals$ is called an \emph{$\cala$-pattern}.  For $M\in \NN$ we denote by $C_M$ the cube at the origin with side length $M  - 1$, i.e.
$$C_M:=\{ x\in \ZZ^d:  0\leq x_j \leq M-1, j=1,\ldots, d\}.$$
The set of all $C_M$, $M\in \NN$, is denoted by $\cC$, 
and a  pattern with  $Q(P) \in \cC$ is called a \emph{cube pattern} or a \emph{box pattern}.
The set of box patterns $P$
with $Q(P) = C_M$ is denoted by $\calpM$. 
For a pattern $P$ and $Q\in \cals$
with $Q \subset Q(P)$ we define the restriction $P\cap Q$ of $P$ to
$Q$ in the obvious way by $P\cap Q : Q \longrightarrow \cala$,
$x\mapsto P(x)$. For a pattern $P$ and $\gamma\in \Gamma$ we define
the shifted map $\gamma P$ by $ \gamma P : \gamma Q(P) \longrightarrow
\cala, \gamma (y) \mapsto P(y)$. On the set of all patterns we define
an equivalence relation by $P\sim P'$ if and only if there exists a
$\gamma\in\Gamma$ with $\gamma P = P'$. For a cube pattern  $P$ and an arbitrary pattern  $P'$ we define
the number of occurrences of the pattern $P$ in $P'$ by
\begin{equation*}
  \sharp^\Gamma_P P' :=
  \sharp\Bigl\{x\in Q(P') :
P'\cap \bigl(x +  Q(P)\bigr)\sim P\}.
\end{equation*}

Given a set $Q \subset \ZZ^d$ we denote by $V^\partial_Q \subset Q$ 
the inner vertex boundary of $Q$, i.e.~the set of those vertices contained in $Q$
which have a neighbour in the complement $\ZZ^d\setminus  Q $. A sequence $(Q_l)_{l\in\NN}$ of
finite subsets of $\ZZ^d$ is called a \emph{van Hove sequence} in
$\ZZ^{d}$ if  $
  \lim_{l  \to \infty} \frac{|V^\partial_{Q_l}|}{|Q_l|}=0. $

\begin{dfn}
  \label{boundaryterm}
  A map $b \colon \cals \longrightarrow
  [0,\infty)$ is called a \emph{boundary term} if $b(Q) = b( t+ Q)$
  for all $t\in \ZZ^d$ and $Q\in \cals$, $\lim_{j\to \infty}
  |Q_j|^{-1} b (Q_j) =0$ for any van Hove sequence $(Q_j)$, and there
  exists $D>0$ with $b(Q) \leq D |Q|$ for all $Q\in \cals$.
\end{dfn}

\begin{dfn}
  \label{function}
  Let $(X,\|\cdot\|)$ be a Banach space and $F \colon
  \cals\longrightarrow X$ be given.   

  (a) The function $F$ is said to be \emph{almost-additive} if there
  exists a boundary term $b$ such that
  \begin{equation*}
    \left\| F ( \cup_{k=1}^m Q_k ) - \sum_{k=1}^m F(Q_k)\right\| \leq \sum_{k=1}^m b
    (Q_k)
  \end{equation*}
  for all $m\in\mathbb{N}$ and all pairwise disjoint sets $Q_k\in\cals$,
  $k=1,\ldots,m$.

  (b) Let $\colouring :\ZZ^d \longrightarrow \cala$ be a colouring.
  The function $F$ is said to be \emph{$\Gamma$-$\colouring$-invariant} if
  \begin{equation*}
    F( Q) = F (\gamma Q)
  \end{equation*}
  whenever $\gamma\in \Gamma $  and $Q\in\cals$ obey $\gamma (\colouring
  \cap Q) = \colouring \cap (\gamma Q)$.   In this case there exists a
  function $\widetilde{F}$ on the cubes $\cC$ with values in $X$ such that
  \begin{equation*}
    F( \gamma Q) = \widetilde{F} \Bigl(\gamma^{-1}\bigl(\colouring\cap (\gamma  Q)\bigr)\Bigr)
  \end{equation*}
  for cubes $Q\in\cC $ and  $\gamma\in \Gamma$.

  (c) The function $F$ is said to be \emph{bounded} if there exists a
  finite constant $C>0$ such that
  \begin{equation*}
    \|F(Q)\| \leq C |Q|
  \end{equation*}
  for all $Q\in \cals$.
\end{dfn}

\begin{thm} \label{ergodicthm}   Let $\cala$ be a finite set, $\colouring \colon
  \ZZ^{d}\longrightarrow \cala$ an $\cala$-colouring and
  $(X,\|\cdot\|)$ a Banach space. Let $(Q_j)_{j\in\mathbb{N}}$ be a van Hove
  sequence such that for every pattern  $P$ the frequency $\nu_P =
  \lim_{j\to \infty} |Q_j|^{-1} \sharp^\Gamma_P (\colouring \cap Q_{j})$
  exists.       Let $F : \cals
  \longrightarrow X$ be a $\Gamma$-$\colouring$-invariant, almost-additive
  bounded function. Then the limits
  \begin{equation*}
    \overline{F} := \lim_{j\to\infty} \frac{F(Q_j)}{|Q_{j}|}    = \lim_{M\to\infty}
    \sum_{P \in \calpM} \nu_P \frac{\widetilde{F}(P)}{|C_M|}
  \end{equation*}
  exist in the topology of $(X,\|\cdot\|)$ and are equal.
\end{thm}

\begin{rem}  (a)  The theorem is proven in \cite{LenzMV} for $\Gamma = \Gammat$.  The  proof carries over to  give the result for $\Gamma = \Gammaf$ as well.

(b) We have explicit bounds on speed of convergence in terms of speed of convergence of the frequencies. For details we refer to \cite{LenzMV}.

\end{rem}

 The previous result does not require the context of an ergodic action. Instead existence of the frequencies is sufficient.  Of course, existence of frequencies follows for ergodic actions. This is discussed next. 
Let  $(\Omega,\PP)$ be a probability space such that $\Gamma$ 
acts ergodically on $(\Omega,\PP)$.  A random $\cala$-colouring is a map
$$\colouring : \Omega\longrightarrow \bigotimes_{\ZZ^d} \cala \;\:\mbox{with}\;\:\;
\colouring(\gamma (\omega))_{ \gamma^{-1} x} = \colouring(\omega)_{x}$$
for all $\gamma\in \Gamma$ and $x\in
\ZZ^d$.

\begin{lem}  
Let $(Q_j)$ be an arbitrary van Hove sequence.  Then, for almost every
  $\omega\in \Omega$ the frequency $\nu_P = \lim_{j\to \infty}
  |Q_j|^{-1} \sharp^\Gamma_P (\colouring (\omega) \cap Q_{j})$ exists and is independent of $\omega$ for every
 cube pattern $P$.
\end{lem}
\begin{proof}
For a fixed pattern $P$ the frequency exists for almost every $\omega$
by a standard ergodic theorem. As there are only countably many $P$ 
the statement follows.
\end{proof}

The random version of Theorem~\ref{ergodicthm} then reads:
\begin{thm} \label{ergodictheorem-random}  
 Let $\cala$ be a finite set, $(\colouring_\omega)_{\omega\in\Omega}$ be a random $\cala$-colouring and
  $(X,\|\cdot\|)$ a Banach space.  Let $(Q_j)_{j\in\mathbb{N}}$ be a van Hove
  sequence. 
  Let $F_\omega : \cals
  \longrightarrow X$ be a family of $\Gamma$-$\colouring_\omega$-invariant, almost-additive
  bounded functions which is homogeneous, i.e.\ $F_{\gamma(\omega)}(\gamma(Q))=F_\omega(Q)$ for all $\gamma\in\Gamma,Q\in\cals$.
  Then, for almost every $\omega\in\Omega$ the limits
  \begin{equation*}
    \overline{F}_\omega := \lim_{j\to\infty} \frac{F_\omega(Q_j)}{|Q_{j}|}    = \lim_{M\to\infty}
    \sum_{P \in \calpM} \nu_P \frac{\widetilde{F}(P)}{|C_M|}
  \end{equation*}
  exist in the topology of $(X,\|\cdot\|)$ and are equal. In particular, $\overline{F}_\omega$ is almost surely independent of $\Omega$.
\end{thm}
\begin{proof}
Almost sure existence of the limit is a direct consequence of the
previous  lemma and the first theorem of this section.
 In fact, this theorem gives the  explicit formula for the limit in terms of the function $\widetilde{F}$. This shows that the limit does not depend on $\omega$ almost surely.
\end{proof}

\begin{rem}  The theorem is  similar in appearance to the "usual" ergodic theorems. Let us therefore point out the  differences: First of all, the result is valid for functions taking values in a Banach space. This gives quite some additional freedom. In fact,  this freedom  will allows us in the next sections  to conclude uniform (in the energy) convergence of the integrated density of states compared to the usual pointwise  (in the energy) results obtained from e.g. subadditve ergodic theorems. 

Moreover, the (proof of the) theorem gives an explicit  description of elements in the probability space for  which the limit exists.  These elements   turn out to be exactly the   typical elements with respect to the randomness viz the elements having frequencies. 

Finally, let us mention that one can even obtain explicit error bounds on speed of convergence in terms of speed of convergence of the frequencies (see \cite{LenzMV}. 

\end{rem}

\section{Operators on Combinatorial Graphs}\label{s-comb-graphs}
In this section we introduce  finite (hopping) range equivariant operators for graph-like structures over $\ZZ^d$. We then use the result of the previous section to  obtain existence of the integrated density of states. 
Finally, we have a closer look at three instances of this situation. 

\medskip

We consider situations in which a fixed finite dimensional Hilbert space is attached to each vertex in $\ZZ^d$. In order to model this we need some more notation. 
Let $\calh$ be a fixed Hilbert space with dimension $\dim (\calh)< 
\infty$ and norm $\|\cdot\|$. Then, 
\begin{equation*} 
  \elzh:=\{ u \colon \ZZ^d \longrightarrow \calh : \sum_{x\in \ZZ^d} 
  \|u(x)\|^2<\infty\} 
\end{equation*} 
is a Hilbert space. The \emph{support} of $u\in \elzh$ is the set 
of $x\in\ZZ^d$ with $u(x) \neq 0$. 
 
For $x\in \ZZ^d$, we define the natural projection $p_x :\elzh 
\longrightarrow \calh$, $u \mapsto p_x (u) := u(x)$. Let $i_x \colon 
\calh \longrightarrow \elzh$ be the adjoint of $p_{x}$. Similarly, for 
a subset $Q\subset \ZZ^d$ we define $\ell^2 (Q,\calh)$ to be 
the subspace of $\elzh$ consisting of elements supported in $Q$.  The 
projection of $ \elzh$ on $\ell^2 (Q,\calh)$ is denoted by 
$p_Q$ and its adjoint by $i_Q$. 
 
The operators and functions we are interested in are specified in the 
next two definitions. 
 
 \begin{dfn}\label{operator}  
  Let $\cala$ be a finite set, $\colouring\colon\ZZ^d\longrightarrow 
  \cala$ a colouring and $H \colon \elzh \longrightarrow\elzh$ a 
  selfadjoint operator. 
 
  (a) The operator $H$ is said to be of \emph{finite range} if there 
  exists a length $R_{\mathit{fr}}>0$ such that $p_y H i_x =0$, 
  whenever $x,y\in \ZZ^d$ have distance bigger than $R_{\mathit{fr}}$. 
 
  (b) The operator $H$ is said to be \emph{$\colouring$-invariant} if 
  there exists a length $R_{\mathit{inv}}\in \NN$ such that $p_y H i_x 
  = p_{\gamma y} H i_{\gamma x}$ for all $x,y\in \ZZ^d$ and $\gamma \in \Gamma$ obeying 
  \begin{equation*} 
    \gamma \Bigl(\colouring \cap \bigl(C_{R_{\mathit{inv}}} (x)\cup 
    C_{R_{\mathit{inv}}} (y)\bigr) \Bigr)    
    = \colouring\cap \bigl (C_{R_{\mathit{inv}}} (\gamma x)\cup 
    C_{R_{\mathit{inv}}} (\gamma y)\bigr).  
  \end{equation*} 
 
\end{dfn} 
 
For a  given colouring $\colouring$, a finite-range, $\colouring$-invariant operator $H$ is fully determined 
by specifying finitely many $\dim(\calh) \times \dim(\calh)$ matrices $p_y H 
i_x$. In particular, such operators $H$ are bounded.  
 
\begin{dfn}  
  Let $\lr$ be the Banach space of right-continuous, bounded functions 
  equipped with the supremum norm.  For a selfadjoint operator $A$ on 
  a finite-dimensional Hilbert space we define its \emph{cumulative 
    eigenvalue counting function} $n(A) \in 
  \lr$ by setting  
  \begin{equation*} 
    n(A) (\lambda):=\sharp\{\text{eigenvalues of $A$ not exceeding 
    $\lambda$}\} 
  \end{equation*} 
  for all $\lambda\in\mathbb{R}$, where each eigenvalue is counted 
  according to its multiplicity. 
\end{dfn}

\begin{thm}\label{ids}  Let $\colouring\colon\ZZ^d\longrightarrow \cala$ be a colouring and  $(Q_j)_{j\in\mathbb{N}}$ a van Hove sequence along which the 
  frequencies $\nu_P$ of all patterns $P\in \calpM$ exist.  Let 
  $H:\elzh\longrightarrow\elzh$ be a selfadjoint,    $\colouring$-invariant finite-range operator.   Then, there exists a unique probability measure $\mu_H$ 
  on $\RR$ with distribution function $N_H$ such that $\frac{1}{|Q_j|} 
  n( p_{Q_j} H i_{Q_j})$ converges to $N_H$ with respect to the 
  supremum norm as $j\to\infty$.  
\end{thm}

A \textbf{proof} for this theorem can be found in  \cite{LenzMV}).  Here, we only sketch the idea.  We consider the Banach space $\lr$  of all right continuous bounded real-valued functions on the real line with the supremum norm.  The operator $A$ then gives rise to a map from $\cals$ to $\lr$ viz $Q \mapsto  n( A_Q)$, where $A_Q $ denotes the restriction of $A$ to $Q$. By assumption on $A$ this map satisfies the assumptions of Theorem \ref{ergodicthm}. Hence, the desired averages exist by that theorem.

\medskip

Linear algebra and the uniform convergence just established can be used to derive the following two corollaries.

\begin{cor}\label{cor-topf-suppe}  
Assume the situation of the theorem and additionally positivity of the frequency
$\nu_P$ for any pattern $P$ occurring in $\colouring$. Then the spectrum of $H$ 
  is the topological support of $\mu_H$. 
\end{cor}

\begin{rem} The assumption on positivity of frequencies is necessary in order to obtain this result.  Consider e.g.  $\mathit{id}$ on 
  $\ell^2(\ZZ)$ and perform a rank one perturbation $B = \langle 
  \delta_0, \cdot\rangle \delta_0$ at the origin. Then, the IDS of 
  $\mathit{id}$ and of $\mathit{id} + B$ coincide, but their spectra 
  do not. 
\end{rem}

\begin{cor}\label{cor-topf-suppe-kompakt}
 Assume the situation of the previous corollary.   Then the following 
  assertions for $\lambda \in \RR$  are equivalent: 
  \begin{itemize} 
  \item[(i)] $\lambda$ is a point of discontinuity of $N_H$, 
  \item[(ii)] there exists a compactly supported eigenfunction of $H$ 
  corresponding to $\lambda$.   
  \end{itemize}
\end{cor}

\begin{rem}  For random Schr\"odinger operators on $\ZZ^d$ the statement of Theorem \ref{ids}  can be obtained from  continuity of the IDS. (More precisely, continuity of the IDS combined with the well established weak convergence of the eigenvalue counting measures gives  uniform convergence of the IDS).  Thus, the statement of Theorem \ref{ids} is particularly interesing for situations in which this continuity is not valid.  This non-continuity arises due to local geometric structures
 as discussed in the previous corollary.  Specific examples where discontinuites of the IDS exist  are discussed in  Subsection \ref{comb-percolation}. 
\end{rem}

\subsection{Periodic Operators}\label{s-comb-graphs-results}
 The setting just described can easily be applied to periodic operators over $\ZZ^d$. In fact, the additional Hilbert space $\calh$  gives quite some freedom to  consider situations which are only similar to $\ZZ^d$. This is discussed next.

\medskip

Let $G$ be a graph with a countable set of vertices (which we again 
denote by $G$) on which $\ZZ^d$ acts isometrically, freely and cocompactly. 
 Let us denote by $\fundom\subset G$ a 
$\ZZ^d$-fundamental domain. Thus $\fundom$ contains exactly one 
element of each $\ZZ^d$-orbit in $G$.  By the cocompactness 
assumption, $\fundom$ is finite.  This implies in particular that 
the vertex degree of $G$ is uniformly bounded.  From now on the 
fundamental domain $\fundom$ will be assumed fixed.

\begin{rem}
A simple example of such a graph is $\ZZ^d$ with the natural action of the group $ (N\ZZ)^d$ for $N\in \NN$ fixed.
Another example would be the 
Cayley graph $G=\operatorname{Cay}(\mathcal{G},S)$ of a direct product group 
$\mathcal{G}= \ZZ^d\otimes F$, where $F$ is any finite group and $S$ is 
a finite, symmetric set of generators for $\mathcal{G}$. Here  the action of $\ZZ^d$ on $G$
is induced by the (obvious) action of $\ZZ^d$ on itself. Note that even for trivial $F$ we obtain infinitely many different graphs,
namely the Cayley graphs of $\ZZ^d$. 
\end{rem}

We now turn to operators acting on $\ell^2(G)$ and $\elzh$.  Let $A 
\colon \ell^2(G) \to\ell^2(G) $ be a selfadjoint linear operator 
satisfying the following two conditions:

\begin{itemize}
\item $A(x,y)= A(\gamma  x,\gamma  y)$ for all   $ x,y \in G, \gamma \in \ZZ^d$.  (Covariance)
\item  There exists $\rho>0$ with $A(x,y)=0$ whenever the graph distance of $x$ and $y$ exceeds $\rho$. (Finite range). 
\end{itemize}

The setting developed so far looks different from the setting discussed in the previous section. To make the connection we proceed as follows.  Let $\cala$ be a set consisting of one element and let $\colouring$ be the trivial colouring.  Set $\cH := \ell^2(\fundom)$, then $\dim(\cH)=|\fundom|$. We can now define a 
unitary operator $U \colon \elzh \to\ell^2(G)$ in the following way: 
For a $\psi \in \elzh$ and $\gamma\in\ZZ^d$ write $\psi(\gamma) 
=\sum_{i\in \fundom} \psi_i(\gamma) \delta_i$, where $(\delta_i)_{i\in\fundom}$ 
is the standard orthonormal basis of $\ell^2(\fundom)$.  Then, the coefficients $\psi_i (\gamma)$ are uniquely determined. We set $( U \psi) (x):= \psi_i (\gamma)$ where $i\in \fundom$ and $\gamma \in \ZZ^d$ are the unique elements such that $x= \gamma i$. 
 Then, $H = U^* A U$ can easily be seen to be a $\colouring$-invariant operator of finite range.  Moreover, the frequencies of all patterns (occurring in $\colouring$) are positive (and in fact equal to $1$). 
Thus, all the results of the  previous section apply to $H$.  They can  then  be used to  infer the obvious analogues for the operator $A$.

\subsection{Set of Visible Points}\label{visible}
The set of visible points in $\ZZ^d$ is a prominent example (and 
counterexample) in number theory and aperiodic order \cite{BaakeMP-00,Pleasants}. 
In particular its diffraction theory has been well studied. Still, it 
seems that the corresponding nearest-neighbour hopping model had not 
received attention until \cite{LenzMV}.   Here, we shortly discuss the result from there. 

\medskip

The set $\vispt$ of visible points in $\ZZ^d$ consists of the origin 
and all $x\neq 0$ in $\ZZ^d$ with 
\begin{equation*} 
  \{t x : 0< t <1\}\cap \ZZ^d =\emptyset. 
\end{equation*} 
Thus,  $x\neq 0$ belongs to $\vispt$, if and only if the greatest common 
divisor of its coordinates is $1$. The obvious interpretation is that 
such an $x$ can be seen by an observer standing at the origin. This 
gives the name to this set. The characteristic function 
\begin{equation*}  
  \colouring := \chi_\vispt \colon \ZZ^d\longrightarrow \cala:=\{0,1\} 
\end{equation*}  
of $\vispt$ provides a colouring. While $\vispt$ is very regular in many 
respects, it has arbitrarily large holes.  In particular, existence of 
the frequencies $\nu_P$ does not hold along arbitrary van Hove 
sequences. However, as was shown in \cite{Pleasants} (see \cite{BaakeMP-00} for special 
cases as well), the frequencies exist and can be calculated explicitly 
for sequences of cubes centred at the origin.  Moreover, the 
frequencies of all patterns which occur are strictly positive. 
 
Thus, all abstract results discussed  above  are valid for 
$\chi_\vispt$-invariant operators of finite range. One relevant such 
operator is the adjacency operator $A_\vispt$. We finish this 
section by defining this operator: Points $x=(x_1,\ldots,x_d)$ and $y 
= (y_1,\ldots, y_d)$ in $\ZZ^d$ are said to be neighbours, written as 
$x\sim y$, whenever 
\begin{equation*}  
  \sum_{j=1}^d |x_j - y_j|=1. 
\end{equation*} 
Then, $A_\vispt \colon \ell^2 (\ZZ^d) \longrightarrow \ell^2 
(\ZZ^d)$ is defined by 
\begin{equation*} 
  (A_\vispt  u) (x) := \chi_\vispt (x) \sum_{y\sim x: y\in \vispt} 
  u(y) 
\end{equation*}  
for all $x\in\ZZ^{d}$ and all $u\in\ell^2 (\ZZ^d)$.

\subsection{Percolation on Combinatorial Graphs}\label{comb-percolation}
In this section we add some randomness to our model. 
Thus we obtain random operators which are generated by a percolation process on the underlying graph.
Hamilton operators on percolation subgraphs of combinatorial graphs
have been considered in the literature in theoretical physics \cite{deGennesLM-59a,KirkpatrickE-72,ChayesCFST-86},
computational physics  \cite{KantelhardtB-02} (and references therein), and mathematical physics
\cite{BiskupK-01a,KloppN-03,Veselic-05a,Veselic-05b,KirschM-06,MuellerS-07,AntunovicV-08b,AntunovicV-c}.

\medskip

Choose $\Gamma =\ZZ^d$. We start with the deterministic part. 
Fix a finite range selfadjoint operator  
$A\colon\ell^2(\ZZ^d)\to\ell^2(\ZZ^d)$ which is invariant under the trivial colouring where every element of $\ZZ^d$ has the same colour. Thus, $A$ is $\ZZ^d$ periodic. 

To define the random part, let $(\Omega,\PP)$ be a probability space 
and $\tau_\gamma\colon \Omega\to\Omega, \gamma\in\Gamma$, an ergodic 
family of measure preserving transformations.  Furthermore, let 
$\cala$ be an arbitrary finite subset of $\RR \cup \{+\infty\}$ and 
$(\omega,x)\mapsto V(\omega,x)\in \cala$ a random field which is 
invariant under the transformations $\tau_\gamma, \gamma\in \Gamma$. 
More precisely, for all $\gamma \in \Gamma$, $\omega \in \Omega$ and 
$x\in G_d$ we require $V(\tau_\gamma\omega,x)=V(\omega,\gamma x)$. 
Next we define random subsets of $\ZZ^d$ and $\ell^2(\ZZ^d)$ induced by the 
random field $V$.  For each $\omega\in \Omega$ define the subset of 
vertices $G_\omega:= \{x \in \ZZ^d: V(\omega,x) <\infty \}$, the natural 
projection operator $p_\omega\colon\ell^2(G_d)\to \ell^2(G_\omega)$ and 
its adjoint $i_\omega\colon\ell^2(G_\omega)\to \ell^2(G_d)$. 
This gives rise to the random    Hamiltonian 
\begin{equation*} 
  H_\omega := A_\omega + V_\omega, \quad \domain(H_\omega):= \ell^2(G_\omega).
\end{equation*} 
Here, the hopping part is given by
\begin{equation*} 
  A_\omega := p_\omega A \, i_\omega, \quad \domain(A_\omega):= \ell^2(G_\omega). 
\end{equation*} 
and $V_\omega$ is defined by $V_\omega:=p_\omega V(\omega, \cdot) 
\, i_\omega \colon \ell^2(G_\omega)\to\ell^2(G_\omega)$. We extend this operator to $\ell^2 (\ZZ^d)$ by setting it equal to zero on the complement of $G_\omega$. The extension will be denoted by $H_\omega$ as well.   For such operators the existence of the IDS as a pointwise limit 
has been established in \cite{Veselic-05a}, and its continuity properties  
have been analysed in \cite{Veselic-05b}.  Here, we discuss  uniform existence of the integrated density of states by fitting these operators into the framework presented  above. 
For each $\omega \in \Omega$ we define a colouring by 
\begin{equation*} 
  \colouring_\omega\colon \ZZ^d \to \cala , 
  \quad \colouring_\omega(x):= V(\omega,x). 
\end{equation*} 
Then, the $(H_\omega)$ are a $\colouring_\omega$-invariant family of operators in the sense of Section \ref{s-ET}. Moreover, it is not hard to see that  
for any pattern $P \colon Q(P) \to \cala, Q(P)\in \cals$, the 
frequency $\nu_P$ of $P$ in $\colouring_\omega$ exists almost surely along the van Hove sequence 
of boxes $C_j,j \in\NN$.  We can then find a set of full measure in  $\Omega$ for which the frequencies of all patterns exist and the frequencies of the occurring patterns are positive. 
For such an $\omega$ we can then apply  Theorem \ref{ids} and its two corollaries.

\medskip

Let us close this section by pointing out two  situations  to which the 
results  presented here can be easily extended: \\
(1)\quad  The operator $A$ can be allowed to be $(N\ZZ)^d$ periodic. In fact,  one can do a similar analysis for models  based on the graphs introduced in Section \ref{s-comb-graphs-results}. \\ 
(2)\quad  Instead of 
\emph{site}-percolation Hamiltonians one can consider  Hamiltonians on 
\emph{bond}-percolation graphs. For the 
construction of the IDS for such operators, see \cite{KirschM-06}.

Let us mention that the asymptotics of the IDS at spectral edges for various percolation 
models has been analysed in \cite{BiskupK-01a,KloppN-03,KirschM-06,MuellerS-07,AntunovicV-08b,AntunovicV-c}.

\section{Operators on Metric Graphs}\label{s-metric-graphs}
The typical (differential) operators of interest on metric graphs are unbounded,
such that their counting functions are right-continuous, but unbounded.
Subtraction of the counting function of a reference operator yields bounded functions again:
spectral shift functions.
Thus we can apply the results of Section~\ref{s-ET} to the (bounded) spectral shift functions,
which in turn yields results for the counting functions.

\subsection{General Results}\label{s-metric-graphs-general}
We define a metric graph $G_d$ over $\ZZ^d$ in the following way. Let $e_j$, $j=1,\ldots,d$ be the standard basis of the real $d$-dimensional space $\RR^d$. 
Each edge $e\in E_d$ is determined by $\init(e)=x,\term(e)=x+e_j$ for some $x\in V_d=\ZZ^d,j\in{1,\ldots,d}$.
We define the metric graph $G_d$ by identifying each edge $e$ of the combinatorial graph with the interval
$e=[x,x + e_j]$, which in turn can be identified canonically with the interval $(0,1)$.
This procedure induces an orientation on our graph. However, it turns out that
all relevant quantities are independent of the choice of orientation.

We will also need to consider finite subgraphs $G$ of
$G_d$ . By a subgraph we mean a subset of the edges of $G_d$
together with all adjacent vertices. All functions we consider live on
the topological space $G_d$ (as a subspace of $\RR^d$) or subgraphs of it.

\medskip

The operators we are interested in will be defined on the Hilbert space
\[
L^2(E_d):= \bigoplus_{e\in E_d } L^2(e)
\]
and their domains of definition will be subspaces of
\[
W^{2,2}(E_d):= \bigoplus_{e\in E_d } W^{2,2}(e),
\]
where $ W^{2,2}(e)$ is the usual Sobolev space of $L^2(e)$ functions
whose (weak) derivatives up to order two are in $L^2(e)$ as well. 
The restriction of $f\in W^{2,2}(E_d) $ to an edge $e$ is denoted by $f_e$.
  For an edge $e=[\init(e),\term(e)]=[x,x+e_j]$ and $g$ in $W^{2,2}(e)$ the
boundary values of $g$
$$ g(\init(e)):=\lim_{t\searrow 0} g (x + t e_j), \;\:
g(\term(e)):=\lim_{t\nearrow 1} g(\init(e) + t e_j)$$  and the boundary
values of $g'$
\[
g'(\init(e)):=\lim_{\epsilon \searrow 0} \frac{g(x+\epsilon e_j)-g(x)}{\epsilon} 
\quad \text{ and } \quad 
g'(\term(e)):= \lim_{\epsilon \searrow 0} \frac{g(\term(e))-g(\term(e) -\epsilon e_j)}{- \epsilon}
\] 
exist by standard Sobolev type theorems. 
For  $f\in W^{2,2}(E_d)$ and each vertex $x$ we gather the boundary values
of $f_e (x)$ over all edges $e$ adjacent to $x$ in a vector
$f(x)$. Similarly, we gather the boundary values of $f_e' (x)$ over all
edges $e$ adjacent to $x$ in a vector $f' (x)$.

Given the boundary values of functions, we can define
boundary conditions following \cite{KostrykinS-99b,Harmer-00}.
A single-vertex boundary condition at $x\in V$ 
is a choice of subspace $L_x$ of $\CC^{4d}$ with
dimension $2d$ such that
 $$ 
 \eta((v,v'),(w,w')):=\langle v',w\rangle - \langle v, w'\rangle
 $$ 
vanishes for all $(v,v'), (w,w')\in L_x$.  An $f\in W^{2,2}(E_d)$ is said to
satisfy the single-vertex boundary condition $L_x$ at $x$ if $(f(x),f'(x))$ belongs
to $L_x$.
A field  of single-vertex boundary conditions $ L:=\{L_x : x\in V_d\}$ will be called 
boundary condition. Given such a field, we obtain a selfadjoint realization $\Delta_L$ 
of the Laplacian $\Delta$ on $L^2 (E_d)$ by choosing the domain
 $$ 
 \domain(\Delta_L) :=\{f\in W^{2,2} (E_d) :\forall x : (f(x),f'(x))\in L_x\}
 $$
and by letting $\Delta_L$ act on $f_e$ as $-f_e''$.
This way all so-called graph-local boundary conditions can be realised, i.e.\ those which 
relate boundary values at the same vertex only.
This includes Dirichlet boundary
conditions with subspace $L^D$ consisting of all those $(v,v')$ with $v=0$, Neumann conditions with
subspace $L^N$ consisting of all those $(v,v')$
with $v'=0$, and Kirchhoff (also known as free) boundary conditions $L^K$ consisting
of all $(v,v')$ with $v$ having all components equal and $v'$ having the sum over its components equal to $0$.

Note that a metric graph together with a field of boundary conditions is sometimes called a ``quantum graph'',
although this means that on a fixed quantum graph there exist only well-defined operators of fixed order (2 in this case).

Everything discussed so far including existence of limits of functions at
the vertices and the notions of boundary condition extends in the
obvious way to subgraphs.  Moreover, for a subgraph $G$ of $G_d$ with edge set $E$, we
write $W^{2,2} (E):=\oplus_{e\in E} W^{2,2} (e)$. The number of edges of
a finite subgraph $G$ of $G_d$ is denoted by $|E|$.

\medskip

 In order to define random operators we need some further data
including a probability space $(\Omega,\PP)$ and an action of (a
subgroup of) the automorphism group of $G_d$ on $\Omega$ and maps $L$,
$V$ from $\Omega$ into the space of boundary conditions and potentials,
respectively.
 As discussed at the beginning, for  us two groups will be relevant, the full
automorphism group $\Gammaf$ and the group $\Gammat$ of translations
by $\ZZ^d$.  We fix one of them, denote it by $\Gamma$   and assume that it acts ergodically on
$(\Omega,\PP)$ via measure preserving transformations. To simplify the
notation 
we identify $\gamma\in \Gamma$ with
the associated measure preserving transformation.

\bigskip

Let us describe the type of random operators we  consider in this section:

 \begin{ass}\label{(S)}
Let $(\Omega,\PP)$ be a probability space and $\Gamma\in \{\Gammaf,\Gammat\}$ a group 
acting ergodically on $(\Omega,\PP)$. 
Let $\cB$ be a finite subset of
$L^\infty(0,1)$ and $\cL$ a finite set of boundary conditions. A random potential is a map
\begin{equation}
\label{e-equiV}
V : \Omega\longrightarrow \bigotimes_{e\in E_d} \cB \;\:\mbox{with}\;\:\;
V(\gamma (\omega))_{\gamma (e)} = V(\omega)_{e}
\end{equation}
for all $\gamma\in \Gamma$ and $e\in E_d$. A random boundary condition is a map 
\begin{equation}
\label{e-equiL}
L : \Omega\longrightarrow \bigotimes_{v\in V_d} \cL,\;\:\mbox{with}\;\:\; L(\gamma (\omega))(\gamma (x)) = L(\omega)(x)
\end{equation}
for all $\gamma\in \Gamma$ and $v\in V_d$.

A family of random operators $(H_\omega)$
on $L^2 (E_d)$ can be defined with domain of definition
$$ \domain(H_\omega ):=\{f\in W^{2,2} (E_d) : (f(x),f' (x)) \in L(\omega)(x)\;\:\mbox{ for all $x\in V_d$}\}$$
acting by
$$ (H_\omega  f)(e) := - f_e'' + V(\omega)_e f_e$$
for each edge $e$.    These are selfadjoint lower bounded operators.
A particularly simple random operator $(H_\omega)$ is given by the 
pure Laplacian $-\Delta_D$ with the domain 
\begin{align*}
\domain(\Delta)
:=\{f \in W^{2,2} (E_Q)\mid & \forall x \in V: (f(x),f'(x))\in L^D\}
\end{align*}
and $-\Delta=H_\omega$ with $V(\omega)\equiv 0$.
 \end{ass}

We assume throughout the section that Assumption \ref{(S)} holds, 
and for this reason do not repeat it in every lemma.

\begin{rem}  While $\Gammaf$ is not commutative it is a natural
object to deal with. In particular, let us note that the Laplacian
without potential and boundary conditions in all vertices identical to
Kirchhoff conditions is invariant under $\Gammaf$.
\end{rem}

We will need to consider restrictions of our operators to finite subgraphs.  These are finite subgraphs associated to finite subsets of $\ZZ^d$. The cardinality of a finite subset $Q$ of $\ZZ^d$ is denoted by $|Q|$.
We introduce the set of edges 
\[
E_Q:=\{e \in E_d\mid \init(e) \in Q\}
\]
and the set of vertices
\[
V_Q:=\{v \in V_d\mid v \text{ adjacent to $e$ for some } e \in E_Q  \}.
\]
The subgraph $(V_Q,E_Q)$ will be denoted by $G_Q$.
Note that $V_Q \supset Q$. 
The set $V^i_Q$ of inner
vertices of $G_Q$ is then given by those vertices of $G_Q$ all of
whose adjacent edges (in $G_d$) are contained in $G_Q$.  The set of inner edges
$E^i_Q$ of $G_Q$ is given by those edges whose both endpoints are inner.
The vertices of $G_Q$ which are not inner are called boundary
vertices. The set of all boundary vertices is denoted by
$V^\partial_Q$.  Similarly, the set of edges which are not inner is
denoted by $E^\partial_Q$.

\smallskip

The restriction $H_\omega^Q$  of the  random operator $H_\omega$  to  $G_Q$ has domain given by
\begin{align*}
\domain(H_\omega^Q):=
\{f \in W^{2,2} (E_Q)\mid & \forall x \in V^i_Q : (f(x),f'(x))\in L(\omega)(x), \\
 &\forall x \in V^\partial_Q : (f(x),f'(x)) \in L^D\}.
\end{align*}
This operator is again selfadjoint, lower bounded, and has purely discrete spectrum. Let us enumerate the eigenvalues
of $H_\omega^Q$ in ascending order
$$\lambda_1(H_\omega^Q) < \lambda_2(H_\omega^Q) \le \lambda_3(H_\omega^Q) \le \dots$$
and counting multiplicities.
Then, the eigenvalue counting function $n_\omega^Q$ on $\RR$ defined by
\[
n_\omega^Q (\lambda) := \sharp \{n \in \NN \mid \lambda_n(H_\omega^Q) \le \lambda\}
\]
is monotone increasing and right continuous, i.e.~a distribution function,
which is associated to a pure point measure, $\mu_\omega^Q$.  Denote by
$$
N_\omega^Q(\lambda) := \frac{1}{ |E_Q|} n_\omega^Q (\lambda)
$$
the volume-scaled version of $n_\omega^Q(\lambda)$ and note that
$ |E_Q| = d |Q|$
as the edge to vertex ratio in the graph $(V_d,E_d)$ is equal to $d$.

  For a finite subgraph $H$ of $G_d$ let
$\chi_{H }$ be the multiplication operator by the characteristic function
of  $H$. Denote the trace on the operators on $L^2(E_d)$ by  $\Tr[\cdot]$.

\medskip

As discussed above, a random Schr\"odinger $(H_\omega)$
as well as the pure Laplacian $\Delta_D$ can be restricted to the subgraphs $G_Q$ induced by
finite sets $Q$ of $\ZZ^d$. This yields the operator $H_\omega^Q$ and
$\Delta_D^Q$ with spectral counting functions $n_\omega^Q$ and $n_D^Q$
respectively. Now, $n_D^Q$ decomposes as a direct sum of
operators. Thus, denoting the eigenvalue counting function of the
negative Dirichlet Laplacian on $]0,1[$ by $n_D (\lambda)$ (similarly as $N_\omega^Q$ above), we have
$n_D^Q = |E_Q| n_D = d |Q| n_D$.  The associated spectral shift
function is given as
$$
\xi_\omega^Q (\lambda) := n_\omega^Q (\lambda)- d \, |Q| \, n_D(\lambda) = d \, |Q| \,
\big(N_\omega^Q (\lambda)-n_D(\lambda)\big).
$$
The crucial point is that
$\xi_\omega$ falls into the framework of almost additive $F$
introduced above. This is shown in the following lemma from \cite{GruberLV-07}.

\begin{lem}
Let $(\rightcont,\|\cdot\|_\infty)$ be the Banach space of right
  continuous bounded functions on $\RR$. Then, for each $\omega\in
  \Omega$ the function $\xi_\omega : \cals \longrightarrow \rightcont$,
  $Q\mapsto \xi_\omega^Q$, is a bounded, $\colouring(\omega)$ invariant
  almost additive function.
\end{lem}
\begin{rem}  The need to use a spectral shift function, i.e. the
difference between $n_\omega$ and $n_D$, in the above lemma comes
exclusively from the boundedness requirement. Note that 
the notion of boundedness depends on the norm of the considered Banach space,
cf.~also Lemma \ref{Rtilde} below. 
\end{rem}

The key result is now the following proposition.

\begin{prp} \label{conv} 
There is a bounded right continuous function $\Xi\colon \RR \to \RR$
such that for a given van Hove sequence $(Q_l)$ for  almost every $\omega \in \Omega$ the uniform convergence
\[
\lim_{l \to\infty}  \Big\|\frac{\xi_\omega^{Q_l}}{|E_{Q_l}|}-\Xi \Big\|_\infty =0\]
holds.
\end{prp}
\begin{proof}
Given the lemma above, the proposition is a direct consequence of Theorem~\ref{ergodictheorem-random}.
\end{proof}

By identifying the limit above and adding it to the IDS of the Dirichlet operator (see \cite{GruberLV-07}) we obtain a Shubin-Pastur
type formula:
\begin{thm}
\label{t-uniformIDS}
Let $Q$ be a finite subset of $\ZZ^d$.
Then, the function $N=N_H$ defined by
 \begin{equation} \label{e-IDS}
 N(\lambda):= \frac{1}{|E_Q|}\int_\Omega   \Tr\left[\chi_{G_Q} \chi_{]-\infty,\lambda]}(H_\omega) \right]  d \PP(\omega)
 \end{equation}
does not depend on the choice of $Q$,  is the distribution function of a measure $\mu= \mu_H$,  and
for any
van Hove sequence $(Q_l)$ in $\ZZ^d$
 \[
 \lim_{l \to\infty} \|N_\omega^{Q_l} -N\|_\infty =0
 \] 
for almost every $\omega \in \Omega$. In particular, for almost every
$\omega\in\Omega$, $N_\omega^{Q_l} (\lambda)$ converges as $l\to\infty$ pointwise to $N(\lambda)$ for every
$\lambda\in\RR$.
\end{thm}

While the definition of the IDS involves an ergodic theorem, there are other spectral features
of $H_\omega$ whose almost sure independence of $\omega$ uses only the ergodicity of the group action.
Prominent examples are the spectrum $\sigma(H_\omega)$ and its subsets 
$\sigma_{pp}(H_\omega)$, $\sigma_{sc}(H_\omega)$, $\sigma_{ac}(H_\omega)$, $\sigma_{disc}(H_\omega)$, $\sigma_{ess}(H_\omega)$ 
according to the spectral type.  In fact, by applying the general framework of \cite{LenzPV-07} we immediately infer the following theorem. 
\begin{thm} There exist subsets of the real line $\Sigma$, $\Sigma_{pp}$, $\Sigma_{sc}$, $\Sigma_{ac}$, $\Sigma_{disc}$, $\Sigma_{ess}$ 
and an $\Omega'\subset\Omega$ of full measure such that $\sigma(H_\omega)=\Sigma$ 
 and $\sigma_\bullet(H_\omega)=\Sigma_\bullet$ for all these spectral types $\bullet \in \{pp,sc,ac,disc,ess\}$
and all $\omega \in\Omega'$.
\end{thm}

\medskip
 
In the following  we discuss three types of percolation models.
These models are based solely on random boundary conditions. The potential of the operators is identically equal to zero.
 Unlike in the percolation models on combinatorial graphs ``deleted'' edges are not
removed completely from the graph but only cut off by Dirichlet boundary
conditions. The reason is that removing edges would mean removing
infinite dimensional subspaces from our Hilbert space. This would result 
in a spectral distribution function which is not comparable to the one of
the Laplacian with the concerned edge included.  

Before giving details we would like to emphasize the following: The
examples below include cases in which the graphs  contain infinitely many 
finite components giving rise to compactly supported eigenfunctions.
In particular, the integrated density of states has a dense set of discontinuities. 
In fact, in the subcritical phase the IDS is a step function, albeit with dense jumps.
However, despite all these jumps our result
on uniform convergence does hold!

\subsection{Site Percolation on Metric Graphs}\label{metric-site-percolation}
The percolation process is defined by the following procedure:
toss a (possibly biased) coin at each vertex and
-- according to the outcome -- put either a Dirichlet or a Kirchhoff
boundary condition on this vertex. Do this at every vertex independently of all the others.
 To be more precise, let $p\in (0,1)$ and $q=1-p$ be given. 
Let $\cala :=\{L^D, L^K\}$ and the probability measure $\nu:= p \delta_{L^K}+ (1-p)\delta_{L^D}$ on $\cala$ be given. 
Define $\Omega$ as the cartesian product space $\times_{x \in V_d} \cala$ with product measure
$\PP:=\otimes_{x\in V_d} \nu$. Let $L$ be the stochastic process 
with coordinate maps 
$L(\omega)(x) := \omega (x)$. These data yield a family of random operators 
$-\Delta_\omega :=H_\omega$  acting like the free Laplacian with domain given by
$$D( \Delta_\omega)=\{f\in W^{2,2} (E) : (f(x),f'(x))\in
\omega(x)  \ \forall x \in V_d\}.$$
Intuitively, placing a Dirichlet boundary condition at a vertex means ``removing'' it from the
metric graph. The $2d$ formerly adjacent edges do not ``communicate''
any longer through the vertex.
A fundamental result of percolation theory tells us that 
for sufficiently small values of $p$ the percolation graph consists entirely of finite components almost 
surely.
For these values of $p$ our Laplace operators decouple
completely into sums of operators of the form $-\Delta^G$ for finite
connected subgraphs $G$ of $G_d$. Here, $\Delta^G$ acts like the free
Laplacian and has  Dirichlet boundary conditions on its deleted vertices (boundary vertices) and 
Kirchhoff boundary conditions in its vertices which have not been deleted by the percolation process 
(interior vertices). 
We introduce an equivalence relation on the set of connected subgraphs of 
$G_d$ with a finite number of edges by setting $G^1 \sim G^2$ iff there 
exists 
a $\gamma \in \Gamma$ such that $ \gamma G^1 = G^2$. For such an 
equivalence class $\cG$ we define 
$n^\cG$  as the eigenvalue counting function of $-\Delta^G$ for some $G \in 
\cG$, 
and set $N^\cG= \frac{n^\cG}{|E_G|}$.
Defining the density  $\nu_\cG$
of an equivalence class of  finite subgraphs of $G_d$ within the 
configuration $\omega$ in the obvious way, 
we obtain
as integrated density of states for the family $H_\omega$
$$ N = \sum_{\cG} \nu_\cG N^\cG,$$ where the sum runs over all equivalence 
classes $\cG$ of finite
connected subgraphs of $G_d$. 
Thus, the integrated density of states is a pure
point measure in this case with many jumps. More interestingly, all
these jumps remain present (even if their height is diminished) when we
start increasing $p$. This yields models in which the operators are
not given as a direct sum of finite graph operators but still have
lots of jumps in their integrated density of states.
Related phenomena for combinatorial Laplacians
have been studied e.g.~in \cite{ChayesCFST-86,Veselic-05b}.

\subsection{Edge Percolation on Metric Graphs}\label{metric-edge-percolation}
The basic idea is to
decide for each edge independently whether Dirichlet boundary conditions
are put on both ends or not. All other boundary conditions are
Kirchhoff type. The problem when defining this edge percolation model is that our stochastic processes
are indexed by vertices rather than edges. We thus have to relate
edges to vertices. This is done by going to each vertex and then tossing
a (biased) coin for each $j=1,\ldots, d$ to decide how to deal with
the edge $[x,x+e_j]$.

More precisely: Let $p_0\in (0,1)$ and
$p_1= 1 - p_0$ be given.  Let $\cala$ consist of all maps $S$ from
$\{1,\dots,d\}$ to $\{0,1\}$.  Put a probability measure $\nu$ on
$\cala$ by associating the value $\prod_{j=1}^d p_{S(j)}$ to the element
$S$. Now, $\Omega$ is the cartesian product space $\times_{x \in V_d} \cala$  with product
measure $\PP:=\otimes_{x\in V_d} \nu$. To each $\omega\in \Omega$ we
associate the operator $-\Delta_\omega =H_\omega$ which acts like the
free Laplacian and has boundary conditions as follows: The edge $e
=[x,x+e_j]$ has Dirichlet boundary conditions on both ends if the random variable associated to the vertex $x$
has the $j$-th component equal to $1$.  Otherwise the boundary condition is chosen to be
Kirchhoff. Here, again the operator decouples completely into 
operators on finite clusters for small	
enough values of $p_0$.

\subsection{Site-Edge Percolation on Metric Graphs}\label{metric-site-edge-percolation}
Similar to the previous two models one can consider a percolation process indexed by pairs $(x,e)$
of adjacent vertices and edges. As in the last model consider a colouring
$\cala$ consisting of all maps $S$ from $\{1,-1,2,-2\dots,d, -d\}$ to $\{0,1\}$.
The probability space and measure are defined similarly as before.
Each $\omega$ gives rise to a Laplace operator with the following boundary conditions:
if the $-j$-th component of the random variable associated to the vertex  $x$ has the value one,
then the edge $[x-e_j,x]$ is decoupled from $x$ by a Dirichlet boundary condition.
If the $j$-th component of the same random variable has value one then the edge 
$[x,x+e_j]$ is decoupled from $x$ by a Dirichlet boundary condition. 
Conversely, those components of the random variable which are zero
correspond to Kirchhoff boundary conditions.

\subsection{Operators with Magnetic Fields}\label{magnetic}
Our set-up is general enough to include magnetic fields as well. 
To this end, let $G$ be a metric graph and $L$ a choice of boundary conditions
as in Section \ref{s-metric-graphs}. The most general symmetric
first order perturbation of $-\frac{\dd^2}{\dd t^2} $ on an edge $e\in
E$ is, up to zeroth order terms, given by 
\[ H(a)_e := -\left(\frac{\dd}{\dd t}-\imath a_e\right)^2 \]
for arbitrary real valued $a_e\in C^1(\bar e)$, where $\bar e$ is the
closure of the edge $e$, i.e.\ identified with the closed interval
$[0,1]$. The selfadjoint realization of $H(a)$ corresponding to $L$ is then given
by the domain
\[ \domain(H_L(a)) = \{f\in W^{2,2}(E): \forall x\in V: (f(x),f'(x)-\imath
(af)(x)) \in L_x \} \]
as the usual partial integration argument shows;
i.e.\ one has to specify mixed Dirichlet and (magnetic) Neumann boundary
conditions as expected.

Now, simple calculations show that this operator is unitarily equivalent
to a nonmagnetic operator with boundary conditions $\tilde L=uL$, where
$(uL)_x:=u_xL_x$ for $x\in V$, and $u_x$ is defined as follows:
Set $\varphi_e(t)=\int_0^t a_e(s)\,\dd s$ for each edge $e$.
Then $u_x$ is a diagonal matrix in $\CC^{2\deg x}$, and the entry belonging to an edge $e$
incident to $x$ is $e^{-\imath \varphi(\init(e))}=1$ if $x=\init(e)$ and $e^{-\imath \varphi(\term(e))}$ if $x=\term(e)$.


Finally, let us note that the above implies that our
results for Schr\"odinger operators with random (or fixed) boundary conditions lead to the same
results for magnetic Schr\"odinger operators with
random (or fixed) magnetic fields, specified through the phases
$\varphi_e(\term(e))$ at the endpoints.

\subsection{Decorated Graphs over $\ZZ^d$}\label{metric-decorated}
Just as in Sections~\ref{s-comb-graphs-results} and~\ref{comb-percolation}, we may consider more general graphs:
Let $G$ be a metric graph such that the group $\Gamma=\ZZ^d$ acts isometrically, freely and cocompactly on $G$. 
This metric space can be viewed as the standard graph over $\ZZ^d$, 
but decorated with compact graphs corresponding to a fundamental domain $\fundom$ for the $\Gamma$-action.
Assume that there is an ergodic action $\tau_\gamma, \gamma \in \Gamma$ of the group 
by measure preserving transformations on the probability space $(\Omega,\PP)$. 
Let a random potential $V$ and a random set of boundary conditions $L$ be given which take on only finitely many values
and satisfy the compatibility conditions \eqref{e-equiV} and \eqref{e-equiL}, where now the edge set $E_d$ is replaced by the edge set $E(G)$ of the graph $G$. This gives rise to a random Schr\"odinger operator $(H_\omega)$
defined on the domain $\domain(H_\omega ) \subset W^{2,2} (E(G))$ which is determined by boundary conditions $L_\omega$. 
For a finite cube $Q\subset \ZZ^d$ set $\fundom(Q):= \bigcup_{\gamma \in Q} \gamma\fundom$
and denote by $(H_\omega^Q)$  the restriction of $(H_\omega)$ to $L^2(\fundom(Q))$. Now we can again 
define the functions $n_\omega^Q (\lambda), N_\omega^Q (\lambda)$ and $\xi_\omega^Q (\lambda) $
as in Section \ref{s-metric-graphs-general}. For these objects the results formulated in Section 
\ref{s-metric-graphs-general} hold true.


\section{Random Graph Metrics}\label{s-rndlen}
In this section we discuss Laplace operators on metric graphs which have random edge lengths.
In this case, even the shift functions will be unbounded. 
Nevertheless, there are two ways
in which we can apply our results from Section~\ref{s-ET}:
Restrict our attention to a bounded energy interval on which the shift functions are uniformly 
bounded; 
or use a different global norm, adjusted to the common growth rate of these functions.
In fact, the IDS will have the same growth as the shift function in this case, which
is why there is no loss in working with the IDS directly.

Let us provide, resp.\ recall, the  notation used in this context.

Let $0<l_- \le l_+ < \infty$ and $\tilde\cala$ a finite subset of $[l_-,l_+]$.
Let $l_e \colon \Omega \to\tilde\cala$ 
be a collection of random variables indexed by the edges $e \in E$.
Similarly as in Section~\ref{s-metric-graphs-general}, for each $\omega \in \Omega$
a metric graph $G_d=G_d(\omega)$ is given by the combinatorial graph $(V_d,E_d)$ and by identifying
each edge $e\in E_d$ with the interval $\big(0,l_e(\omega)\big)$.
Recall that there is a group $\Gamma$ acting by isometries on the set of vertices as well as on the set of edges. 
The same group acts ergodically by measure preserving transformations on the probability space $\Omega$.
We assume that the random family of edge lengths obeys the transformation rule
\[
l_{\gamma(e)} (\gamma\omega) = l_e (\omega).
\]
In particular, the distribution of the random variable $l_e$ is independent of $e$.

Now we define a colouring of the vertex set. Let $\cala= \tilde\cala^d$. 
For each $\omega \in \Omega$ define a colouring map  by $\tilde\colouring(\omega) \colon\ZZ\to \cala$ by
$\tilde\colouring(\omega)(v)= \big (l_{e_1}(\omega), \dots,l_{e_d}(\omega)\big)$.
Here $e_j$ is the edge with $\init (e_j)=v=(v_1,\dots,v_d)$ 
and $\term (e_j)=(v_1,\dots,v_{j-1}, v_j +1, v_{j +1}, \dots, v_d)\in\ZZ^d$.

The value $l_e(\omega) $ denotes the length of the edge $e$ of the metric graph $G_d(\omega)$ 
in the configuration $\omega$. 
For each configuration $\omega$ we thus obtain a Laplace operator $H_\omega$ on $G_d(\omega)$.
If $Q\subset \ZZ^d$ is a cube in the vertex set of $G_d(\omega)$ 
we introduce the set of edges
 $E_Q=\{e \in E_d\mid \init(e) \in Q\}$ 
and the set of vertices
$V_Q=\{v \in V_d\mid v \text{ adjacent to $e$ for some } e \in E_Q  \}$
as above.
Note that the sets $V_Q$ and $E_Q$ are independent of $\omega$.
The subgraph $(V_Q,E_Q)$ with length function $l(\omega)|_{E_Q}$ will be denoted by $G_Q:=G_Q(\omega)$. In this situation we consider
again the restriction (with Dirichlet b.c.) of $H_\omega$ to $G_Q(\omega)$ and denote it by 
$H_\omega^Q$, while its eigenvalue counting function is denoted by $n_\omega^Q \colon \RR \to\RR$.

Next we show an analogue of Lemma 21 in \cite{GruberLV-07}.
\begin{lem}
\label{l-conditions-satisfied}
Let $a <b\in\RR$ be given. Let $(\rightcont,\|\cdot\|_\infty)$ be the Banach space of right
  continuous bounded functions on $[a,b]$ equipped with the supremum norm. Then, for each $\omega\in
  \Omega$ the function $n_\omega : \cals \to \rightcont$,
  $Q\mapsto n_\omega^Q$, is a  $\tilde\colouring(\omega)$-invariant, almost additive, bounded function.
\end{lem}
\begin{proof} Almost additivity and invariance are shown exactly as in Lemma 21 in \cite{GruberLV-07}.
We show now that $n_\omega$ is bounded in the sense of Definition \ref{boundaryterm}.
For the Dirichlet Laplace operator on $[0,l]$ the eigenvalue counting function
$n_d(l,\lambda) \colon \RR \to\RR$ is given by
\[
n_d(l,\lambda) = \left\lfloor \frac{l}{\pi} \sqrt{\lambda}\right\rfloor. 
\]
It follows that 
\begin{equation}
\label{e-difference}
n_d(l,\lambda) \le \frac{\sqrt{\lambda}}{\pi} \; l + 1
\end{equation}
Now we can argue again as in Lemma 21 in \cite{GruberLV-07}. First we change all 
boundary conditions of $H_\omega^Q$ to Dirichlet ones. This corresponds to a perturbation operator of rank at most $2 |E_Q|$. Consequently, it contributes an error term  of at most $2 |E_Q|= 2d|Q|$.
Subsequently we use for each decoupled edge the estimate
\eqref{e-difference}. It follows that for $\lambda \in [a,b]$
\begin{equation}
\label{e:boundedness}
\begin{aligned}
|n_\omega^Q(\lambda) |
& \le  2|E_Q| + \sum_{e\in E_Q} N(l_e,\lambda)
\\ &\le  2|E_Q| +    \sum_{e \in E_Q} \left( \frac{\sqrt{\lambda}}{\pi} l_e +1 \right)
\\ & =  3|E_Q| +   \frac{\sqrt{\lambda}}{\pi} \sum_{e \in E_Q} l_e
\\ & \le 3|E_Q| +   \frac{\sqrt{b}}{\pi} |E_Q| l_+
\\ & = |Q|d \left( 3 +   \frac{\sqrt{b}}{\pi} l_+ \right)
\end{aligned}
\end{equation}
\end{proof}

\begin{rem}  The reason why we obtain uniform convergence only on bounded energy intervals
and not on the whole axis is the last step in the proof of Lemma \ref{l-conditions-satisfied}. 
Using the SSF instead of the IDS we would merely obtain a smaller constant, but the same qualitative behaviour.
\end{rem}

The previous Lemma allow us to prove

\begin{prp} \label{p:localy-uniform-convergence} 
There is a right continuous function $N\colon \RR \to \RR$
such that for a given van Hove sequence $(Q_l)$and  for any bounded interval $I \subset \RR$  
the uniform convergence
\[
\lim_{l \to\infty}  \sup_{\lambda \in I}\Big|\frac{n_\omega^{Q_l}}{|E_{Q_l}|}(\lambda)-N(\lambda) \Big|=0
\]
holds for  almost every $\omega \in \Omega$.
\end{prp}
\begin{proof}
By Lemma~\ref{l-conditions-satisfied}, each $n_\omega$ is invariant, almost additive and bounded.
The family $(n_\omega)_{\omega\in\Omega}$ is homogeneous by construction, so that Theorem~\ref{ergodictheorem-random} applies.
\end{proof}

\bigskip

Now we present a variant of the above result. We consider a different Banach space, 
namely
\[
\tilde\rightcont
:= \{ f\colon \RR \to\RR\mid f \text{ is right continuous and } \sup_\RR \big|f(x)/ \sqrt{|x|+1}\big| <\infty  \}
\]
with the norm $\sup_\RR \big|f(x)/ \sqrt{|x|+1}\big|$.
Again we have: 
\begin{lem} \label{Rtilde}
For each $\omega\in \Omega$ the function $n_\omega \colon \cals \to \tilde\rightcont$,
$Q\mapsto n_\omega^Q$, is  $\tilde\colouring(\omega)$-invariant, almost additive and bounded.
\end{lem}
\begin{proof} Almost additivity and invariance are shown as before.
To show boundedness we note that from  \eqref{e:boundedness} it follows that 
\[
\sup_{\lambda \in \RR} |n_\omega^Q(\lambda)/\sqrt{|\lambda|+1}| 
\le \frac{3 |E_Q| }{\sqrt{|\lambda|+1}} +\frac{\sqrt{\lambda}}{\sqrt{|\lambda|+1}} \frac{|E_Q|}{\pi} 
l_+
\le const |E_Q|
\]
Here $const$ is some constant independent of $Q$, $\omega$ and $\lambda$. 
Recall that $|E_Q|= d |Q|$. Thus we have proven boundedness.
\end{proof}
Thus the statement of Proposition \ref{p:localy-uniform-convergence} can be strengthened to
\[
\lim_{l \to\infty}  \sup_{\lambda \in \RR}\Big|\Big(\frac{n_\omega^{Q_l}}{|E_{Q_l}|}(\lambda)-N(\lambda) \Big) \frac{1}{\sqrt{|\lambda|+1}}  \Big|=0
\]
for  almost every $\omega \in \Omega$.

\section{Outlook: Uniform Convergence and Jumps of the IDS}\label{s-jumps}
In this section we provide an outlook beyond the theorems stated in the main text.
This includes a discussion of the relation to other (recent) papers, as well as some open questions
which we plan to address in the future.

\subsection*{Uniform Convergence and Discontinuities of the IDS}
Our goal here  is to   elaborate on the  relationship between the almost sure spectrum of $H_\omega$ and its	 IDS. 

We start by noting two corollaries  of  Theorem~\ref{ids}  for  the setting of combinatorial graphs and of  
Theorem~\ref{t-uniformIDS}  for quantum graphs  respectively. Note that in contrast to Corollaries~\ref{cor-topf-suppe} and~\ref{cor-topf-suppe-kompakt}
we do not require positivity of frequencies, since we are in the random setting now.  We recall the notions of topological support $\supp \mu$ of  $\mu=\mu_H$ and the set
$S_p (\mu):=\{\lambda\in \RR: \mu(\{\lambda\})>0\}$ of atoms of  $\mu$ to the spectrum of 
$H_\omega$. Note that the set of discontinuities of the IDS is precisely  $S_p (\mu)$.

\begin{cor}\label{spectrum}
$\Sigma$ equals the topological support $\supp \mu$ of $\mu$.
\end{cor}

As usual an $f\in\elzh$ is said to be compactly supported if $f(v)=0$ for all but finitely many $v\in V$; 
$f\in W^{2,2} (E_d)$ is said to be compactly supported if $f_e\equiv 0$ for all but finitely many edges $e$.

\begin{cor}\label{jump}
Denote by $\Sigma_{cmp}$ the set of energies $\lambda\in \RR$
such that there exists almost surely a compactly supported square integrable eigenfunction $f_\omega$ with
$H_\omega f_\omega = \lambda f_\omega$. Then
\begin{equation} \label{e-jumps}
S_p(\mu) = \Sigma_{cmp}.
\end{equation}
\end{cor}

\begin{rem} 
\begin{enumerate}[(a)]
\item 
Note that there are many examples where the IDS has discontinuities: 
the free Laplacian (i.e.~the Schr\"odinger operator with identically vanishing potential)
with Dirichlet, Neumann, or Kirchhoff boundary conditions; 
percolation models such as those in Sections~\ref{comb-percolation}, \ref{metric-site-percolation}, \ref{metric-edge-percolation} and \ref{metric-site-edge-percolation}; 
for other 
percolation and tiling Hamiltonians, and quantum graphs, see also \cite{ChayesCFST-86,KlassertLS-03,KostrykinS-04,Veselic-05b}.

\item
If the randomness entering the potential of the operator is sufficiently strong it is natural
to expect a smoothing effect on the IDS. In fact, in \cite{HelmV} for a class of alloy-type random potentials
the Lipschitz-continuity of the IDS was established. 
In \cite{GruberV,GruberHV} we  show for a different class of random potentials how one can 
estimate the modulus of continuity of the IDS. Such estimates are relevant in the context of spectral 
localisation for random Schr\"odinger operators, see the disussion in \cite{GruberHV} in this volume.
\end{enumerate}
\end{rem}

These corollaries provide a criterion for deducing the existence of discontinuities of the IDS
from uniform convergence (w.r.t.~the energy parameter) of the IDS. 
More precisely,
they prove that (A) implies (B'), where
\begin{itemize}
\item[(A)]  $ \displaystyle \lim_{l\to\infty} \| N_\omega^{Q_l} -N_H\|_\infty=0$, \\
i.e~the IDS is uniformly approximated by its finite volume analoga 
\end{itemize}
and
\begin{itemize}
\item[(B')] The IDS has discontinuities precisely at those energies, which are eigenvalues
of $H_\omega$ with compactly supported eigenfunctions almost surely.
\end{itemize}

This line of argument stems originally from \cite{KlassertLS-03}. See \cite{LenzMV,GruberLV-07} for the present context.
It is possible to turn the argument around. More precisely, property
\begin{itemize}
\item[(B)] The positions and the sizes of the jumps of the IDS are approximated by the
analogous data of the finite volume approximands $N_\omega^{Q_l}$.
\end{itemize}
implies already (A).

Ideas of this type have been used in \cite{Eckmann-99,Elek-03,MathaiSY-03}.
More recently, in \cite{LenzV} it was proven that (B), and hence (A), holds for general ergodic, equivariant, selfadjoint, finite hopping range operators on discrete structures. The equivariance of the operator is supposed to hold w.r.t.~an amenable group.

Thus for many models it is possible to pursue two different routes to obtain the same results:
\begin{enumerate}[(i)]
\item Either one first establishes a Banach-space valued ergodic theorem, implying (A), and then deduces (B). This was the choice made in the present work.
\item Or one first proves the statement (B) about jumps, and then concludes (A).
\end{enumerate}

The latter approach has  three advantages:
\begin{itemize}
\item It does not require any finite local complexity property, and thus works even for 
random operators where a single matrix element may assume infinitely many values. (This is 
e.g.~the case for Anderson-percolation Hamiltonians, cf.~\cite{Veselic-05b}.)
\item
It works for models which are equivariant w.r.t.~a non-abelian group, as long as it is amenable.
\item
Furthermore, the equivariance group may be discrete (like $\ZZ^d$) or connected (like $\RR^d$).
\end{itemize}
On the other hand the former approach (i) has the advantage of providing certain
information  beyond that obtained from approach (ii). In particular, it 
allows one to control 
\begin{itemize}
\item
the set of measure zero where the convergence fails as well as
\item
the convergence speed in terms of an error estimate. 
\end{itemize}
In certain situations, such as  the setting of minimal, uniquely ergodic dynamical 
systems, the control of the exceptional set of measure zero implies actually that it is empty (see \cite{LenzS-06})!
Thus convergence holds for all configurations of the randomness 
rather than for almost all only.

\subsection*{Absence of Discontinuities}

Above we discussed characterisations of the positions and sizes of 
jumps of the IDS. This allows in particular to show that several models
have an IDS with (many) discontinuities. Intuitively, the discontinuities are related to two facts, namely that
the models in question 
\begin{itemize}
\item
are not \emph{too random}, 
for instance satisfy a finite local complexity condition, and
\item
do not satisfy the (appropriate) unique continuation property. 
\end{itemize}
On the other hand, for certain random Hamiltonians where even single 
matrix elements have continuous distribution it can be shown that the IDS is 
Lipschitz-continuous. These results go under the name of Wegner estimates, cf.~\cite{Wegner-81},
and are well established for operators on $\ell^2(\ZZ^d)$ and $L^2(\RR^d)$, see e.g.~\cite{KirschM-07,Veselic-07b}
for recent surveys. Meanwhile such bounds have been also established for quantum graphs in
\cite{HelmV,GruberV,GruberHV, LenzPPV-II}.

Furthermore, if a model satisfies an appropriate version of the unique continuation property
the IDS has no jumps, regardless of whether the finite local complexity condition is fulfilled or not 
(cf.\ e.g.\ Proposition~5.2. in \cite{Veselic-05b}). An example of such a random operator is the Anderson model for which the continuity  of the IDS was established in \cite{CraigS-83a,DelyonS-84}.
 
This can be compared nicely to the properties of mixed Anderson-percolation Hamiltonians.
There, due to the dilution of the lattice, the operator does not have the unique continuation property. 
For this model it turns out that the IDS has jumps if and only if the distribution of the matrix elements has atoms 
(apart form the point mass at $\infty$ which corresponds to the deletion of a vertex).

\subsection*{Questions}

Let us formulate two questions which concern Banach-space valued ergodic theorems more 
general than those formulated in the present review:

\begin{itemize}
\item Is it possible to show, using a similar line of argument as in \cite{Lenz-02,LenzMV},
that a  Banach-space valued ergodic theorem holds if the lattice $\ZZ^d$ is replaced by 
a finitely generated, discrete group of polynomial growth?

We plan to address this question in the future.
The hope is that --- since such groups are both amenable and residually finite --- one can use 
a similar covering argument as in \cite{LenzMV}.
\item
Is there a version of a Banach-space valued ergodic theorem
which is applicable both to models with $\ZZ^d$-equivariance as well as models with  
$\RR^d$-equivariance, and which provides a unified treatment of the results in \cite{LenzS-06}
and \cite{LenzMV}? This would mean that one does not need to distinguish between a discrete and a continuous group action.
\end{itemize}
 
\renewcommand{\MR}[1]{} 
 \bibliographystyle{amsalpha}
 \bibliography{uids}

\end{document}